\input harvmac 

\let\includefigures=\iftrue
\includefigures
\input epsf
\def\figin{\epsfcheck\figin}\def\figins{\epsfcheck\figins}
\def\epsfcheck{\ifx\epsfbox\UnDeFiNeD
\message{(NO epsf.tex, FIGURES WILL BE IGNORED)}
\gdef\figin##1{\vskip2in}\gdef\figins##1{\hskip.5in}
\else\message{(FIGURES WILL BE INCLUDED)}%
\gdef\figin##1{##1}\gdef\figins##1{##1}\fi}

\def\IP{\relax{\rm I\kern-.18em P}}

\def\IR{\relax{\rm I\kern-.18em R}}
\def\DefWarn#1{}
\def\figinsert{\goodbreak\midinsert}
\def\ifig#1#2#3{\DefWarn#1\xdef#1{fig.~\the\figno}
\writedef{#1\leftbracket fig.\noexpand~\the\figno}%
\figinsert\figin{\centerline{#3}}\medskip\centerline{\vbox{\baselineskip12pt
\advance\hsize by -1truein\noindent\footnotefont{\bf Fig.~\the\figno:} #2}}
\bigskip\endinsert\global\advance\figno by1}
\else
\def\ifig#1#2#3{\xdef#1{fig.~\the\figno}
\writedef{#1\leftbracket fig.\noexpand~\the\figno}%
\global\advance\figno by1} 
\fi

%
\ifx\epsfbox\UnDeFiNeD\message{(NO epsf.tex, FIGURES WILL BE
IGNORED)}
\def\figin#1{\vskip2in}
\else\message{(FIGURES WILL BE INCLUDED)}\def\figin#1{#1}\fi
\def\ifig#1#2#3{\xdef#1{fig.~\the\figno}
\goodbreak\topinsert\figin{\centerline{#3}}%
\smallskip\centerline{\vbox{\baselineskip12pt
\advance\hsize by -1truein\noindent{\bf Fig.~\the\figno:} #2}}
\bigskip\endinsert\global\advance\figno by1}


\overfullrule=0pt
\def\Title#1#2{\rightline{#1}\ifx\answ\bigans\nopagenumbers\pageno0\vskip1in
\else\pageno1\vskip.8in\fi \centerline{\titlefont #2}\vskip .5in}

\lref\mal{J.~Maldacena, 
``The large N limit of superconformal field theories and supergravity,'' 
Adv.\ Theor.\ Math.\ Phys.\ {\bf 2}, 231 (1998) [Int.\ J.\ Theor.\ Phys.\ {\bf 38}, 1113 (1998)] [hep-th/9711200].}

\lref\GomisVK{
J.~Gomis,
``D-branes, holonomy and M-theory,''
Nucl.\ Phys.\ B {\bf 606}, 3 (2001)
[arXiv:hep-th/0103115].}

\lref\KlebanovHB{
I.~R.~Klebanov and M.~J.~Strassler,
``Supergravity and a confining gauge theory: Duality cascades and  chiSB-resolution of naked singularities,''
JHEP {\bf 0008}, 052 (2000)
[arXiv:hep-th/0007191].}

\lref\AcharyaGB{
B.~S.~Acharya,
``On realising N = 1 super Yang-Mills in M theory,''
arXiv:hep-th/0011089.}

\lref\malnu{J.~M.~Maldacena and C.~Nunez,
``Towards the large N limit of pure N = 1 super Yang Mills,''
Phys.\ Rev.\ Lett.\  {\bf 86}, 588 (2001)
[hep-th/0008001].}

\lref\WittenDF{E.~Witten,
``Constraints On Supersymmetry Breaking,''
Nucl.\ Phys.\ B {\bf 202}, 253 (1982).}

\lref\RedlichDV{
A.~N.~Redlich,
``Parity Violation And Gauge Noninvariance Of The Effective Gauge Field Action In Three-Dimensions,''
Phys.\ Rev.\ D {\bf 29}, 2366 (1984).}

\lref\StromingerUH{
A.~Strominger,
``Superstrings With Torsion,''
Nucl.\ Phys.\ B {\bf 274}, 253 (1986).}

\lref\mnas{
J.~Maldacena and H.~Nastase, 
``The supergravity dual of a theory with dynamical supersymmetry breaking,'' hep-th/0105049. }

\lref\jerome {
J.~P.~Gauntlett, N.~Kim, D.~Martelli and D.~Waldram, 
``Wrapped fivebranes and N = 2 super Yang-Mills theory,'' hep-th/0106117. }

\lref\AtiyahQF{
M.~Atiyah and E.~Witten,
``M-theory dynamics on a manifold of G(2) holonomy,''
arXiv:hep-th/0107177.}

\lref\guksp{S. Gukov and J. Sparks, ``M-Theory on Spin(7) Manifolds:
I'', hep-th/0109025.}%

\lref\VafaWI{
C.~Vafa,
``Superstrings and topological strings at large N,''
arXiv:hep-th/0008142.}

\lref\zaff{F.~Bigazzi, A.~L.~Cotrone and A.~Zaffaroni, 
``N = 2 gauge theories from wrapped five-branes,'' hep-th/0106160.} 

\lref\oht{K.~Ohta, 
``Moduli space of vacua of supersymmetric Chern-Simons theories and type IIB branes,'' JHEP {\bf 9906}, 025 (1999) [hep-th/9904118]. }

\lref\ohtm{B.~Lee, H.~Lee, N.~Ohta and H.~S.~Yang, 
``Maxwell Chern-Simons solitons from type IIB string theory,'' Phys.\ Rev.\ D {\bf 60}, 
106003 (1999) [hep-th/9904181].} 

\lref\PapadopoulosGJ{
G.~Papadopoulos and A.~A.~Tseytlin,
``Complex geometry of conifolds and 5-brane wrapped on 2-sphere,''
Class.\ Quant.\ Grav.\  {\bf 18}, 1333 (2001)
[arXiv:hep-th/0012034].}

\lref\WittenDS{
E.~Witten,
``Supersymmetric index of three-dimensional gauge theory,''
arXiv:hep-th/9903005.}

\lref\KaoGF{
H.~C.~Kao, K.~y.~Lee and T.~Lee,
``The Chern-Simons coefficient in supersymmetric Yang-Mills Chern-Simons theories,''
Phys.\ Lett.\ B {\bf 373}, 94 (1996)
[arXiv:hep-th/9506170].}

\lref\berg{O.~Bergman, A.~Hanany, A.~Karch and B.~Kol, 
``Branes and supersymmetry breaking in 3D gauge theories,'' JHEP {\bf 9910}, 036 (1999) [hep-th/9908075]. }

\lref\GomisAA{
J.~Gomis and J.~G.~Russo,
``D = 2+1 N = 2 Yang-Mills theory from wrapped branes,''
arXiv:hep-th/0109177.}

\lref\Alvarez{
L.~Alvarez-Gaume and E.~Witten,
``Gravitational Anomalies,''
Nucl.\ Phys.\ B {\bf 234}, 269 (1984).}

\lref\kol{O.~Bergman, A.~Hanany, A.~Karch and B.~Kol,
``Branes and supersymmetry breaking in 3D gauge theories,''
JHEP{\bf 9910}, 036 (1999)
[hep-th/9908075].}

\lref\CveticDM{
M.~Cvetic, H.~Lu and C.~N.~Pope,
``Consistent Kaluza-Klein sphere reductions,''
Phys.\ Rev.\ D {\bf 62}, 064028 (2000)
[arXiv:hep-th/0003286].}

\lref\GukovHF{
S.~Gukov and J.~Sparks,
``M-theory on Spin(7) manifolds. I,''
arXiv:hep-th/0109025.}

\lref\japanone{
T. Kitao, K. Ohta and N. Ohta,
``Three-Dimensional Gauge Dynamics from Brane Configurations with
$(p,q)$-Fivebranes'', Nucl. Phys. {\bf B539} (1999) 79-106, [hep-th/9808111];
K.~Ohta,
``Moduli space of vacua of supersymmetric 
Chern-Simons theories and type  IIB branes,''
JHEP{\bf 9906}, 025 (1999)
[hep-th/9904118];
B.~Lee, H.~Lee, N.~Ohta and H.~S.~Yang,
``Maxwell Chern-Simons solitons from type IIB string theory,''
Phys.\ Rev.\ D {\bf 60}, 106003 (1999)
[hep-th/9904181]; T.~Kitao and N.~Ohta,
``Spectrum of Maxwell-Chern-Simons theory 
realized on type IIB brane  configurations,''
Nucl.\ Phys.\ B {\bf 578}, 215 (2000)
[hep-th/9908006];
K.~Ohta,
``Supersymmetric index and s-rule for type IIB branes,''
JHEP{\bf 9910}, 006 (1999)
[hep-th/9908120].}

\lref\kol{O.~Bergman, A.~Hanany, A.~Karch and B.~Kol,
``Branes and supersymmetry breaking in 3D gauge theories,''
JHEP{\bf 9910}, 036 (1999)
[hep-th/9908075].}

\lref\GauntlettUR{
J.~P.~Gauntlett, N.~w.~Kim, D.~Martelli and D.~Waldram,
``Fivebranes wrapped on SLAG three-cycles and related geometry,''
arXiv:hep-th/0110034.}

\lref\SinhaAP{
S.~Sinha and C.~Vafa,
``SO and Sp Chern-Simons at large N,''
arXiv:hep-th/0012136.}

\lref\ChamseddineNM{
A.~H.~Chamseddine and M.~S.~Volkov,
``Non-Abelian BPS monopoles in N = 4 gauged supergravity,''
Phys.\ Rev.\ Lett.\  {\bf 79}, 3343 (1997)
[arXiv:hep-th/9707176].}

\lref\levelrank{S.~G.~Naculich, H.~A.~Riggs and H.~J.~Schnitzer,
``Group-level duality in WZW models and Chern-Simons theory,''
Phys.\ Lett.\ B {\bf 246}, 417 (1990); S.~G.~Naculich and 
H.~J.~Schnitzer, ``Duality relations between $SU(N)_k$ and $SU(k)_N$
WZW models and their braid matrices,'' Phys.\ Lett.\ B {\bf 244},
235 (1990); I.~B.~Frenkel, in: Lie algebras and related topics, Lecture
Notes in Mathematics, Vol. 93, ed. D. Winter (Springer, Berlin,1982), 
p.71; J.~Fuchs and P.~van Diel, ``Some symmetries of quantum dimensions,''
 J.\ Math.\ Phys.\ {\bf 31}, 1220 (1990); A.~Kuniba and T.~N.~Nakanishi,
``Level-rank duality in fusion RSOS models,'' Bombay\ Quant.\ Field\ 
Theory 1990, 344; M.~Camperi, F.~Levstein and G.~Zemba, ``The large 
N limit of Chern-Simons gauge theory,'' Phy.\ Lett.\ B {\bf 247}, 549
(1990)
}

\lref\ChamseddineMC{
A.~H.~Chamseddine and M.~S.~Volkov,
``Non-Abelian solitons in N = 4 gauged supergravity and leading order
string theory,'' 
Phys.\ Rev.\ D {\bf 57}, 6242 (1998)
[arXiv:hep-th/9711181].}

\lref\AtiyahZZ{
M.~Atiyah, J.~Maldacena and C.~Vafa,
``An M-theory flop as a large N duality,''
arXiv:hep-th/0011256.}

\lref\AcharyaMU{
B.~S.~Acharya, J.~P.~Gauntlett and N.~Kim,
``Fivebranes wrapped on associative three-cycles,''
Phys.\ Rev.\ D {\bf 63}, 106003 (2001)
[arXiv:hep-th/0011190].}

\lref\FreedVC{
D.~S.~Freed and E.~Witten,
``Anomalies in string theory with D-branes,''
arXiv:hep-th/9907189.}

\lref\ChamseddineHK{
A.~H.~Chamseddine and M.~S.~Volkov,
``Non-Abelian vacua in D = 5, N = 4 gauged supergravity,''
JHEP {\bf 0104}, 023 (2001)
[arXiv:hep-th/0101202].}

\lref\SchvellingerIB{
M.~Schvellinger and T.~A.~Tran,
``Supergravity duals of gauge field theories from SU(2) x U(1) gauged  supergravity in five dimensions,''
JHEP {\bf 0106}, 025 (2001)
[arXiv:hep-th/0105019].}

\lref\HernandezBH{
R.~Hernandez,
``Branes wrapped on coassociative cycles,''
arXiv:hep-th/0106055.}

\lref\DeserWH{
S.~Deser, R.~Jackiw and S.~Templeton,
``Topologically Massive Gauge Theories,''
Annals Phys.\  {\bf 140}, 372 (1982)
[Erratum-ibid.\  {\bf 185}, 406.1988\ APNYA,281,409 (1982)].}

\lref\WittenNV{E.~Witten,
``Supersymmetric index in four-dimensional gauge theories,''
arXiv:hep-th/0006010.}

\lref\WittenXY{
E.~Witten,
``Baryons and branes in anti de Sitter space,''
JHEP {\bf 9807}, 006 (1998)
[arXiv:hep-th/9805112].}

\Title{\vbox{\baselineskip12pt \hbox{hep-th/0111060}
\hbox{CALT-68-2358 }  \hbox{CITUSC/01-040} }}
{\vbox{
\centerline {On SUSY Breaking And $\chi$SB}
\bigskip
\centerline{From String Duals}
}}
%
%
\centerline{Jaume Gomis }

\bigskip
\centerline{{\it Department of Physics}}
\centerline{{\it California Institute of Technology}}
\centerline{\it Pasadena, CA 91125}

%
%

\def\[{\left [}
\def\]{\right ]}
\def\({\left (}
\def\){\right )}

\vskip .3in

\centerline{\bf Abstract}

We find regular string duals of three dimensional ${\cal N}=1$ SYM
with a Chern-Simons interaction at level $k$ 
for  $SO$ and $Sp$ gauge groups. Using the string dual we exactly
reproduce the conjectured pattern of supersymmetry breaking
proposed by Witten by showing that there is dynamical supersymmetry breaking
 for $k<h/2$ while supersymmetry remains unbroken for $k\geq h/2$,
where $h$ is the dual Coxeter number of the gauge
group. We also find regular string duals of four dimensional
${\cal N}=1$ SYM for $SO$ and $Sp$ gauge groups and exactly reproduce the
expected pattern of chiral symmetry breaking $Z_{2h}\rightarrow Z_2$
by analyzing the symmetries of the string solution.

\smallskip
\noindent
\Date{}


\newsec{Introduction}

The description of asymptotically free, confining gauge
theories via gravity has shed new light on familiar questions regarding the
non-trivial infrared dynamics of gauge theory. Despite the lack of
proper string techniques to study the strong coupling
dynamics of an asymptotically free dual gauge theory, many
expectations from gauge theory have been realized in gravity
duals. Understanding how to deal with  string theory in highly
curved backgrounds in the presence of Ramond-Ramond fields will
eventually  
yield a calculationally reliable solution of the large N limit of gauge theory.

In this paper we find the string duals of minimally supersymmetric
gauge theories in three and four dimensions for $SO$ and $Sp$ gauge
groups and exactly reproduce some of the expected non-trivial infrared
dynamics of these gauge theories. 

In the three dimensional case we generalize the construction of 
Maldacena and
Nastase \mnas\  and construct regular string duals of ${\cal N}=1$ SYM
with a Chern-Simons interaction at level $k$ for $SO$ and $Sp$ gauge
groups. Witten has conjectured 
that supersymmetry is spontaneously broken in this theory when
$k<h/2$, where $h$ is the dual Coxeter number of the gauge group, and has
shown that supersymmetry is unbroken for $k\geq h/2$. Generalizing the
results of \mnas\ we show that the string dual we describe exactly
reproduces the 
conjectured pattern of supersymmetry breaking.

We also construct a regular supergravity solution describing four dimensional
${\cal N}=1$ SYM for $SO$ and $Sp$ gauge groups. Here we generalize
the solution of Maldacena and Nu\~nez \malnu\ to find solutions in which the
expected pattern of chiral symmetry breaking $Z_{2h}\rightarrow Z_2$
in  gauge theory 
is realized in the 
string construction by showing  that an asymptotic symmetry of
the solution that exists in the UV region is broken in the IR region.

An important tool  in finding the string duals of these gauge
theories is the proper treatment of orientifold projections\foot{In a
similar context to the one studied here, reference \SinhaAP\  considered
orientifolds of wrapped branes.}
 when
branes wrap supersymmetric cycles and understanding how to extend the
orientifold 
action in the gravity dual. Precise agreement with field theory
expectations is obtained by taking into account an interesting
subtlety on the proper description of the Wess-Zumino coupling on a
D-brane in models with unoriented open strings which we describe in
section $3$.

The plan of the rest of the paper is as follows. In section $2$ we
briefly summarize the field theory results in \WittenDS\ and 
the solution found in \mnas\ which is dual to $D=3$ SYM for $SU(N)$
gauge group. In section $4$ we make a suitable orientifold construction which
generalizes the gauge groups to $SO$ and $Sp$ and identify the effect
of the orientifold construction in the gravity dual. Following \mnas\
we realize Witten's conjecture about supersymmetry breaking. In
section $4$ we recall  the field theory results on chiral symmetry
breaking in $D=4$ ${\cal N}=1$ SYM and the dual  supergravity solution in 
\malnu\  when the gauge group is $SU(N)$. In section $5$ we find the
string dual for $SO$ and $Sp$ gauge groups and reproduce using the
string dual the expected pattern of chiral symmetry breaking. Finally,
the appendix rederives the BPS equations in \mnas\ from an effective
superpotential method.

\newsec{SUGRA dual to $D=3$ ${\cal N}=1$ SYM with a Chern-Simons Interaction}

A supersymmetric Chern-Simons interaction at level\foot{In this paper
we will take $k>0$.} $k$ can be added to
${\cal N}=1$ SYM with gauge group $G$
\eqn\chernsimons{
L_{CS}={k\over 4\pi}\int \hbox{Tr}\left(A\wedge dA +{2\over
3}A\wedge A \wedge A+\bar{\lambda}\lambda\right).}
Classically, the supersymmetric gauge theory has a mass gap when
$k\neq 0$
\DeserWH . The mass of the gluon and
gluino fields
is of order $g^2k$
and for $k>>1$ one can integrate out the massive
gluinos while retaining a local purely bosonic low energy effective
action. Integrating out a massive fermion at one
loop in a three
dimensional gauge theory induces a
Chern-Simons interaction  whose level is determined by the
sign of the fermion mass 
and the representation  of $G$ under which the fermion transforms
\Alvarez\RedlichDV \foot{The induced level is given by $\delta
k=-{1\over 2}C_R\ \hbox{sign}(m)$, where $C_R$ is the index
of the representation $R$ under which the fermion transforms and $m$ is the 
fermion mass,
i.e. $\hbox{tr}(t^a_Rt^b_R)=2C_R \delta^{ab}$. For the adjoint representation 
$C_R=h$.}. Therefore, in the 
supersymmetric example we are considering, integrating out the  
massive gluino shifts the effective value of $k$ appearing in the
bosonic Chern-Simons term to $k_{eff}=k-h/2$
\KaoGF , where $h$ is the dual Coxeter number of G. 
At very low energies, where the 
Yang-Mills term can be neglected in comparison with the Chern-Simons
term, the effective description of the model is -- for $k>>1$ --
given by bosonic 
Chern-Simons theory with gauge group G at level $k_{eff}$. 
This result was extended $\it{to\ all}$
$k$ in \WittenDS\  where it was shown  that
the low energy effective description of ${\cal N}=1$ SYM with gauge
group $G$ with a  Chern-Simons interaction at level $k$ is given by
bosonic Chern-Simons theory with gauge group $G$ at level 
$k_{eff}=k-h/2$.  

Using this effective description, Witten \WittenDS\ computed the
supersymmetry index \WittenDF\ $\hbox{Tr}(-1)^F$ of three 
dimensional ${\cal N}=1$ SYM with gauge group $G$ with a Chern-Simons
interaction of level $k$. The
number of supersymmetric vacua equal the number of
zero energy states of the effective bosonic Chern-Simons theory at
level $k_{eff}$, which in turn equal the number of
conformal blocks of the corresponding WZW model\foot{To properly
define the index one considers the 
theory on ${\bf R}\times {\bf T}^2$ and the number vacua equals 
the number of  conformal blocks
of the WZW model on a Rieman surface with ${\bf T}^2$
topology. On ${\bf T}^2$, the conformal blocks are in one to one
correspondence with 
the integrable representations of the affine Lie algebra ${\hat G}$ at
level $k_{eff}$. Moreover, all vacua are bosonic or 
fermionic so that $\hbox{Tr}(-1)^F=\pm$ number of conformal blocks.}.
The result is 
\eqn\index{
\hbox{Tr}(-1)^F=\Big\{ 
\matrix{ \neq 0\quad k\geq h/2 &\cr
=0\quad k< h/2& \cr}}
which shows that supersymmetry is unbroken for $k\geq h/2$. 
The supersymmetry index vanishes for $k<h/2$, which is
insufficient to decide 
whether supersymmetry is broken in the infinite volume
limit. However, in \WittenDS\ Witten conjectured 
that dynamical supersymmetry breaking does occur for $k<h/2$.

In section $3$ we generalize the work of Maldacena and Nastase \mnas\
and exactly reproduce  the conjectured pattern of
supersymmetry breaking for SO and Sp gauge groups
using  a string dual
description.
Here we summarize the
value of the dual Coxeter number of the gauge groups for which we find
a 
supergravity dual solution.

\input tables
\bigskip
\begintable
|~~~~$SU(N)$~~~~|~~~~$SO(2N)$~~~~|~~~~$Sp(N)$~~~~|~~~~$SO(2N+1)$~~~~\crthick
~~~$h$~~~|$N$|$2N-2$|$N+1$|$2N-1$\endtable
\centerline{{\bf Table 1}:\ {\ninerm Value of dual Coxeter number for
several gauge groups.}}

\bigskip

\subsec{Supergravity Dual For $G=SU(N)$}

A simple way\foot{For an alternative realization of $SU(N)$ SYM with a
Chern-Simons  term see for example \japanone\kol , where a description of
supersymmetry breaking for $SU(N)$ was given.}
 of realizing $D=3$, ${\cal N}=1$ SU(N) SYM is to study 
Type II string theory on ${\bf R}^{1,2}\times {\bf X}$, where ${\bf
X}$ is 
a seven
dimensional manifold of $G_2$ holonomy and to consider the low energy
effective field theory living on $N$ suitably  wrapped
branes\foot{${\bf X}$ has to be a non-compact space in order for the
flux coming from the wrapped branes to be able 
to spread to infinity.}. There
are two choices that lead to the desired gauge theory. One can either wrap
$N$ Type IIB five-branes on the supersymmetric ${\bf S}^3$ cycle 
\AcharyaMU\mnas\ of
${\bf X}=S({\bf
S}^3)$ -- the spin bundle over ${\bf S}^3$ -- or wrap $N$ Type IIA
D6-branes on the supersymmetric four cycle
\GomisVK\HernandezBH\GukovHF\
 ${\bf Z}$ of ${\bf X} =\Sigma({\bf
Z})$ -- the bundle of anti-self-dual two-forms over ${\bf Z}$ -- where
${\bf Z}$ is ${\bf S}^4$ or ${\bf \IP}^2$\ \foot{Since ${\bf \IP}^2$ is
not a Spin 
manifold, one must turn on 
non-zero magnetic flux 
through ${\bf \IP}^1\in {\bf \IP}^2$
\FreedVC .}. Since the cycle which the
branes wrap is rigid, the familiar massless scalars
describing the motion of the branes in the transverse directions are
absent in the low energy field theory and one just gets pure $SU(N)$
gauge theory 
in three dimensions. In this paper we consider the Type IIB
realization with wrapped five-branes.

In the Type IIB realization a 
 Chern-Simons interaction at level $k$ can be incorporated by turning on 
 three-form flux through ${\bf S}^3\in S({\bf
S}^3)$ which  induces the following  bosonic low energy coupling on the
five-branes
\eqn\coupling{
L_{WZ}\hskip-2pt=\hskip-2pt{1\over 8\pi^2}\int_{R^{1,2}\times S^3} \hskip-28pt B\wedge \hbox{Tr}(F\wedge
F)\hskip-2pt=\hskip-2pt-{1\over 8\pi^2}\int_{R^{1,2}\times S^3}  \hskip-28pt H\wedge
\hbox{Tr}(A\wedge 
dA +{2\over 3} A^3)\hskip-2pt=\hskip-2pt-{k_0\over 4\pi}\int_{R^{1,2}} \hskip-12pt \hbox{Tr}(A\wedge
dA +{2\over 3} A^3)}
which together with its superpartner generates the ${\cal N}=1$
supersymmetric Chern-Simons interaction \chernsimons .

We have been careful in differentiating between the levels $k_0$
and $k$. The reason for this is that the low energy effective field
theory on the branes is $SU(N)$ SYM with a Chern-Simons interaction with bare level
$k_0$. However, massive ${\cal N}=1$ Kaluza-Klein multiplets  arise by
reducing the six dimensional massless vector multiplet on the five-branes over
${\bf S}^3$ and  integrating out the massive Kaluza-Klein
adjoint fermions 
may shift the bare Chern-Simons level $k_0$ if there is a net
number of multiplets with a given sign for their mass\foot{Since
the shift of the level depends 
only on the sign of the fermion mass term, fermions with opposite mass
leave $k_0$ invariant and the net shift of the level
depends only on the net number of fermions with a specific mass term
sign.}. It has been shown\foot{One can start with the topological 
twist corresponding to 
$D=3$ ${\cal N}=2$ SYM, so that the Kaluza-Klein modes come in
${\cal N}=2$  chiral multiplets which have two fermions one with each
sign for the mass and therefore produce no shift of the level. One can
continuously twist the theory to ${\cal N}=1$ by smoothly changing the
gauge field on the normal bundle. To determine the number of unpaired
fermions one computes the number of eigenvalues of the Dirac operator
which change sign under a gauge transformation of the normal bundle,
which is just given by the winding number of the gauge transformation
\WittenDS ,
which is 1.} in \mnas\ that 
 there is a single Kaluza-Klein adjoint massive chiral multiplet which
 does not have 
a partner with opposite mass
so
that the effective theory on the branes after integrating out all
 Kaluza-Klein modes is three dimensional ${\cal N}=1$
 $SU(N)$ SYM with a Chern-Simons interaction at level 
$k=k_0-h/2=k_0-N/2$.

Recently, Maldacena and Nastase \mnas\ have interpreted a solution\foot{In the appendix
we rederive this solution from an effective superpotential for the ansatz.}
found by Chamseddine and Volkov
\ChamseddineHK\ as describing a collection of $N$ five-branes wrapping 
${\bf S}^3\in S({\bf S}^3)$ when $k_0=N$. 
This regular
supersymmetric solution is therefore dual to $D=3$, ${\cal N}=1$ 
$SU(N)$ SYM with a Chern-Simons term at level $k=N/2$. 



The Type IIB supergravity solution \mnas\ obtained by lifting 
\mnas\SchvellingerIB\ the gauged
supergravity solution in \ChamseddineHK\  to ten dimensions
is in the class of
supersymmetric supergravity solutions considered in 
\StromingerUH \foot{Reference \StromingerUH\ considered the case
 $D=6$. The analysis was extended when $D=7$ in  \GauntlettUR.}. 
It is a compactification\foot{Strictly speaking, it is not a
compactification since the internal geometry is non-compact.} of Type
IIB string theory on ${\bf R}^{1,2}\times {\bf M}_7$ with a nontrivial
three-form field strength $H$ and dilaton profile over ${\bf M}_7$, where ${\bf
M}_7$ denotes the 
internal space. The metric of the solution  is given by
\eqn\solun{\eqalign{
ds^2&=ds^2({\bf R}^{1,2})+ N[dr^2+{1\over 4} R^2(r) w_a w_a]+  N
{1\over 4}({\tilde w}_a-A_a)^2\cr
A_a&={1+b(r)\over 2}w_a,}}
where $w_a, a=1,2,3$ parametrize ${\bf S}^3$ and  ${\widetilde w}_a,
a=1,2,3$ are another set of $SU(2)$ left 
invariant one forms parametrizing ${\widetilde{\bf S}}^3$. The radial
function $b(r)$ is monotically increasing and goes from 0 at
$r\rightarrow \infty$ to 1 at $r=0$. On the other hand, $R^2(r)$ behaves
linearly in r near  $r\rightarrow \infty$ while $R^2(r)\simeq r^2$ near $r=0$,
so that the metric inside the square brackets in \solun\ is the flat ${\bf R}^4$
near the origin. Topologically, ${\bf
M}_7$ is asymptotically conical with base ${\bf S}^3\times {\widetilde
{\bf S}}^3$ and in the interior the ${\widetilde
{\bf S}}^3$ stays of finite size while ${\bf S}^3$ becomes
contractible.  The supersymmetry left unbroken by the solution, which
is ${\cal N}=1$ in three dimensions, guarantees that ${\bf M}_7$ is a $G_2$
holonomy manifold with respect to the connection with torsion
$\bar{\nabla}=\nabla +{\pi} H$, where $\nabla$ is the usual Levi-Civita
connection and $H$ is the three-form piercing ${\bf
M}_7$.

The solution has non-trivial H-flux (see eqns. A.5 and A.15 in the
Appendix). The number of five-branes wrapping ${\bf
S}^3 \in S({\bf S}^3)$ is encoded as magnetic flux of
 $H$ through  ${\widetilde{\bf S}}^3$. The solution also has flux
through ${\bf S}^3$, and its asymptotic value at the boundary
$r\rightarrow \infty$, as explained previously, determines the
Chern-Simons level of the holographic field theory dual. 
The fluxes of the solution are\foot{The flux of $H$ through ${\bf S}^3$ 
vanishes at $r=0$ as required for a smooth solution since ${\bf S}^3$
vanishes in the 
interior.} 
\eqn\fluxes{
\int_{{\widetilde{\bf S}}^3}{H \over 2\pi}=\int_{{{\bf
S}}^3}{H_\infty \over 2\pi}=N.} 
Since $k_0=N$, this supergravity solution is dual to $SU(N)$
SYM with $k=N/2$. An external quark is represented by a 
 string stretching from the boundary 
and since the solution is completely nonsingular and without a horizon 
it  exhibits confinement as well as a mass
gap as expected from the field theory analysis \WittenDS\WittenNV .

We note that this duality  realizes the general philosophy advocated in 
\VafaWI\ in which the infrared description of the gauge theory on a
collection of wrapped branes is given by a topologically distinct
background in which branes have been replaced by fluxes. In the
original description the branes wrap ${\bf S}^3\in S({\bf S}^3)$ which
is asymptotically a cone over ${\bf S}^3\times {\widetilde
{\bf S}}^3$, where ${\widetilde
{\bf S}}^3$ is contractible in the full geometry while  ${\bf S}^3$ is
topologically non-trivial. The dual description of the gauge theory is
given by a geometry ${\bf M}_7$ in which the roles of ${\bf S}^3$ and
${\widetilde 
{\bf S}}^3$ are reversed, so that  ${\widetilde
{\bf S}}^3$ is now topologically non-trivial and supports the flux
left behind by the branes once they disappear. In effect the infrared
dynamics of the gauge theory is captured by the ``flopped'' geometry
which the branes wrapped
\AtiyahZZ .

Since the supergravity solution is supersymmetric, this shows that
$SU(N)$ SYM with a Chern-Simons term at level $k=N/2$ 
has unbroken supersymmetry. Maldacena and Nastase showed that one
can use this solution to explore what happens when $k\neq
N/2$. Changing to $k=N/2+n$ -- or equivalently changing to $k_0=N+n$
-- is accomplished by 
wrapping $n=k-N/2$ five-branes\foot{Since the external quark is a string
which can now end on the five-branes, this shows that for $k\neq N/2$
that there is no confinement, in agreement with the field theory
analysis of \WittenDS .}
on ${\widetilde{\bf S}}^3\in {\bf M}_7$,
which adds $n$ units of H-flux through ${\bf S}^3\in {\bf M}_7$ so
that the level is $k=N/2+n$. This can be done reliably for $n<<N$,
when the
backreaction of the $n$ wrapped branes can be safely
ignored. Adding $n>0$ five-branes does 
not break supersymmetry\foot{The $G_2$ structure of ${\bf M}_7$, which
is a three-form, is such that when pulled back to ${\widetilde{\bf
S}}^3$ is proportional to the volume form on  ${\widetilde{\bf
S}}^3$.}, since the wrapped branes preserve the same
supesymmetry as the background \mnas . On the other hand, adding $n<0$
five-branes -- that is $|n|$ anti-branes -- breaks supersymmetry
completely. Therefore, using supergravity one realizes the expected
pattern of supersymmetry breaking  for $G=SU(N)$. For $k\geq N/2$
supersymmetry is unbroken and for $k<N/2$ supersymmetry is
broken. 

For $k>N/2$ the low energy effective description of
the physics is given by the field theory on the $n$ five-branes
wrapped over ${\widetilde{\bf S}}^3$. 
At very low energies, where the Yang-Mills term can be neglected, one
obtains ${\cal N}=1$ $SU(n)$ Chern-Simons theory with level $k=N$.
Integrating out the $SU(n)$ adjoint gluinos and Kaluza-Klein modes result in 
 a negligible shift of the Chern-Simons level (of ${\cal O}(n)$) in
the regime where backreaction can be ignored so that the 
 effective description is given
by $SU(n)$ Chern-Simons theory with level $k=N$
 which is precisely
the level-rank
\levelrank\ 
 dual of what was found in \WittenDS\ which is
$SU(N)$ Chern-Simons theory with $k_{eff}=n$ . In the next 
section we generalize this duality to SO and Sp gauge groups.

\newsec{Supergravity Dual For $G=SO$ and $G=Sp$}


The $SU(N)$ theory is realized by wrapping $N$ D5-branes\foot{In the
previous section we wrote the solution corresponding to wrapped
NS5-branes. One can trivially act with S-duality and write the
corresponding solution for wrapped D5-branes, which is given by
$ds^2(D5)=e^{-\phi_{NS}}ds^2(NS5), C_2^R=-B_2^R$ and $\phi_{D5}=-\phi_{NS}$.}
 on ${\bf S}^3\in
S({\bf S}^3)$. A simple way to generalize the construction to other
gauge groups is to superimpose an orientifold five-plane -- an
{\bf O}5-plane -- on top of  the five-branes. Since the branes are curved this
requires performing an orientifold projection which leaves an {\bf O}5-plane
along ${\bf R}^{1,2}\times {\bf S}^3$, which is the worldvolume of the
five-branes. This can be accomplished by
considering the following projection 
\eqn\proj{
\hbox{Type IIB on}\ \left({\bf R}^{1,2}\times S({\bf
S}^3)\right)/\{1,\Omega\cdot \sigma\},} 
where $\Omega$ is the usual worldsheet parity operator
 and $\sigma$ is
the generator of a $Z_2$ symmetry of  $S({\bf S}^3)$ which we will now
describe. 

The topology of $S({\bf S}^3)$ where the five-branes are
supersymmetrically embedded can be characterized by the following
hypersurface in ${\bf R}^8$ 
\eqn\hyper{
\sum_{i=1}^4 a_i^2-b_i^2=V,}
with $a_i,b_i\in {\bf R}$. The space is asymptotically conical and the
base of the cone is ${\bf S}^3\times {\widetilde {\bf S}}^3$. 
These spheres are parametrized  by the coordinates $a_i$ and
$b_i$ respectively . For $V>0$,  ${\bf S}^3$ is topologically non-trivial while
${\widetilde {\bf S}}^3$ shrinks smoothly in the interior and is therefore
topologically trivial. For $V<0$ the roles of ${\bf S}^3$ and
${\widetilde {\bf S}}^3$ are reversed and now it is ${\widetilde {\bf
S}}^3$ which is topologically non-trivial. The transformation
$V\rightarrow -V$ can then be interpreted as a flop transition 
\AtiyahZZ .

In this description of 
$S({\bf S}^3)$ the relevant $\sigma$ acts by
\eqn\action{
{\sigma}:\Big\{ 
\matrix{ a_i& \rightarrow & a_i& \cr
b_i& \rightarrow & -b_i& \cr}.}
If one considers the geometry near infinity after quotienting by
$\sigma$ one gets a cone with base ${\bf S}^3\times{\widetilde{\bf
S}}^3/Z_2$. Since  ${\widetilde{\bf
S}}^3$ is contractible in the full geometry the $Z_2$ action has fixed points.
Therefore, an {\bf O}5-plane sits at the fixed locus of $\sigma$ -- that
is $b_i=0$ -- which is ${\bf R}^{1,2}\times {\bf S}^3$, which is what
we were after. Therefore, we
have identified a suitable orientifold projection in string theory
that results, at energy scales smaller than the scale of the ${\bf S}^3$
to a three-dimensional ${\cal N}=1$ gauge theory with $SO$ or $Sp$
gauge groups. For future reference we note that when $V<0$, that is in
the flopped geometry, that $\sigma$ is a freely acting involution
since the fixed point set $b_i=0$ is not part of the
manifold and thus there is no orientifold plane.

Three different gauge groups can be realized by placing D5-branes on
top of an {\bf O}5-plane. The three types of orientifold
planes\foot{There is actually a fourth type \WittenXY\ of {\bf O}5-plane, which
yields $Sp(N)$ gauge symmetry. It differs from the familiar {\bf
O}5-plane yielding the same gauge group by a discrete Ramond-Ramond 0-form
$\theta$-angle. These two theories differ only in their
nonperturbative spectrum.}
differ in the choice of discrete fluxes in the geometry transverse to
the orientifold plane, which in this case
is ${\widetilde{\bf RP}}^3={\widetilde{\bf S}}^3/Z_2$.

\input tables
\bigskip
\begintable
|~~~~${\bf O}5^-$~~~~|~~~~${\bf O}5^+$~~~~|~~~~$\widetilde{{\bf O}5}^-$~~~~\crthick
~~~$G$~~~|$SO(2N)$|$Sp(N)$|$SO(2N+1)$\ \endtable
\centerline{{\bf Table 2}:\ Gauge group on  $2N$ five-branes coincident
with an {\bf O}5-plane.} 

\bigskip

We can now construct SUGRA duals to these gauge theories by
implementing the orientifold action we have just described on the
SUGRA solution presented in the previous section . That background was
given by Type IIB string theory on ${\bf R}^{1,2}\times {\bf
M}_7$ where ${\bf
M}_7$ is a $G_2$ holonomy manifold with respect to a connection with
torsion. As discussed in the previous section, topologically ${\bf
M}_7$ is 
asymptotic to  a cone over ${\bf S}^3\times {\widetilde{\bf S}}^3$
just like $S({\bf S}^3)$. However, ${\bf M}_7$ is topologically the flopped
version of $S({\bf S}^3)$.
That is,  in ${\bf M}_7$  ${\bf S}^3$ is contractible while  ${\widetilde{\bf
S}}^3$  is topologically non-trivial. 
Therefore, $\sigma$ acts freely on ${\bf M}_7$ since
${\widetilde{\bf RP}}^3$ never shrinks. 
Thus,  the string dual
to $D=3$ ${\cal N}=1$ SYM 
theories with $SO$ and $Sp$ gauge groups is described by unoriented
closed string theory on the  smooth background ${\bf R}^{1,2}\times {\bf
M}_7/Z_2$, without any open strings
nor orientifold planes.

In the SUGRA dual solution the branes wrapped on ${\bf S}^3$ disappear
and are replaced by flux over ${\widetilde{\bf S}}^3$, which is now
non-contractible. This  still occurs when an {\bf O}5-plane is
added in the gauge theory realization. The only two differences are that 
one must take into account the total five-brane charge  -- that
of the five-branes plus that of 
the {\bf O}5-plane -- which is then converted into flux in the dual theory and
replace ${\widetilde{\bf 
S}}^3$ by ${\widetilde{\bf
RP}}^3$ due to the action of $\sigma$.
The total five-brane charge can be easily computed by remembering that
the gauge groups that we are considering appear by 
projecting the massless open strings on  $N$ branes
together with their images -- so that there are $2N$ branes
in total -- approaching an orientifold plane 
and that the charge of an  ${\bf O}p^{\pm}$ plane is $\pm 32\times
2^{p-9}$. Finally, one must recall that an $\widetilde{\bf O}p^-$ plane has an
extra Dp-brane which can't be moved from the orientifold plane since
it has no image under the orientifold action. 
Therefore, taking into account the charge of the
orientifold
 we see that the total five-brane charge  is given by

\input tables
\bigskip
\begintable
|~~~~${\bf O}5^-$~~~~|~~~~${\bf O}5^+$~~~~
|~~~~${\widetilde {\bf O}5}^-$~~~~\crthick   
~~~$G$~~~|$SO(2N)$|$Sp(N)$|$SO(2N+1)$\ \crthick 
~~~$Q$~~~|$2N-2$|$2N+2$|$2N-1$\endtable
\centerline{{\bf Table 3}:\ Five-brane charge of a configuration of $2N$
branes on top of an {\bf O}5-plane.}
\bigskip

\noindent

As described above the corresponding SUGRA solution
 is found from
that in \mnas\  by replacing
everywhere in the solution $N\rightarrow Q$ and ${\widetilde{\bf
S}}^3\rightarrow  {\widetilde{\bf
RP}}^3$. Therefore, the fluxes of the supergravity solution are given by
\eqn\fluxes{
\int_{{\widetilde{\bf S}}^3}{H \over 2\pi}=\int_{{{\bf
S}}^3}{H_\infty \over 2\pi}=Q.}
Since there is non-zero flux of $H$ through ${\bf
S}^3$ this solution describes $D=3$ ${\cal N}=1$ SYM with a
non-vanishing Chern-Simons interaction. In order to identify the
correct value of the level of the Chern-Simons interaction, one must
take into account the following subtlety\foot{A similar discussion
appeared recently in 
\AtiyahQF }.

The Wess-Zumino term 
\eqn\wesszumino{
L_{WZ}={1\over 8\pi^2}\int B\wedge \hbox{Tr}F\wedge F}
is the familiar coupling in the
worldvolume theory of  $N$
D5-branes with gauge group $SU(N)$. This coupling implies that 
a D1-brane dissolved inside  N D5-branes is equivalent
to a single instanton in the $SU(N)$ gauge theory on the
five-branes, so that the induced D1-string charge is the same as the instanton
number in the $SU(N)$ gauge theory. Is this coupling modified when one
considers the worldvolume theory of D-branes with unoriented open
strings?
First let's consider the case where the gauge theory on the brane is
$SO(2N)$\foot{The discussion that follows trivially extends when
the gauge group is $SO(2N+1)$.}. When the five-branes approach the ${\bf
O}5^-$-plane the gauge symmetry is enhanced from $SU(N)$ to $SO(2N)$. Since a
single instanton in $SU(N)$ corresponds to a unit of D1-string charge,
in order to determine the correct normalization of the Wess-Zumino
coupling for an $SO(2N)$ gauge theory, one must determine the $SO(2N)$
instanton number of a $SU(N)$ gauge field configuration when it is
embedded in $SO(2N)$. It is straightforward\foot{One can embed a
$SU(N)$ instanton in $SO(2N)$ by turning on a $SU(2)$ instanton in
$SU(N)$ via the embedding given by $SU(N-2)\times SU(2) \subset
SO(2N-4)\times SO(4)$. The minimal $SO(2N)$ instanton is then obtained
by embedding the $SU(2)\subset SU(N)$ instanton in an $SU(2)$ subgroup of
$SO(4)$, which clearly 
has instanton number one in $SO(2N)$.} 
 to show that an instanton
number one gauge field in $SU(N)$ has instanton number one when
embedded in $SO(2N)$, which in turn corresponds a single unit of induced
D1-string charge. Therefore the correct coupling is given by
\eqn\wesszumSoP{
L_{WZ}^{SO}=L_{WZ}.}

On the other hand,  when the five-branes approach the ${\bf
O}5^+$-plane the gauge symmetry is enhanced from $SU(N)$ to
$Sp(N)$. Here an instanton number one gauge field in $SU(N)$ has
instanton number ${\it two}$ when embedded in $Sp(N)$. This follows by
embedding an $SU(2)$ instanton inside $SU(N)$ into $Sp(N)$ via the splitting
$SU(N-2)\times SU(2) \subset Sp(N-2)\times Sp(2)$. If we denote by $t$ an
$SU(2)$ gauge field configuration, then the minimal embedding of
$SU(2)$ inside $Sp(2)$ is given by the following $Sp(2)$ Lie algebra
valued matrix
\eqn\embeddP{
T_{Sp}=\pmatrix{t& 0\cr
0 & -t^T}.} 
It then follows that the instanton number of
$T_{Sp}$ is twice that of $t$. Therefore, the induced
D1-string charge on the D5-brane worldvolume equals ${\it twice}$  the 
instanton number of $Sp(N)$ so that the correct Wess-Zumino coupling
is given by
\eqn\wesszumSoP{
L_{WZ}^{Sp}={1\over 2}L_{WZ}.}

Taking into account this subtlety for $Sp(N)$ one finds that the 
the flux of $H$ over ${\bf S}^3$ in Table 3 induces  a bare
Chern-Simons interaction at level $k_0=h$ (see Table 1) on the
worldvolume of the 
wrapped branes for all gauge groups. As explained in \mnas , $k_0$   is shifted by the
presence of an unpaired massive Kaluza-Klein fermion, which is not
projected out by 
the presence of the {\bf O}5-plane but transforms now in the adjoint
representation of $G$. Integrating this fermion out shifts $k_0$ to
$k=k_0-h/2=h/2$.
Therefore, we have
generalized the results of of \mnas\ to find the string dual  of 
three dimensional ${\cal N}=1$ 
SYM with gauge group G and a Chern-Simons
coupling at level $k=h/2$  and have seen that supersymmetry is
unbroken. Regularity of the solution and the lack of a horizon show
that the dual gauge theory confines and has a mass gap as expected
from the field theory analysis of \WittenDS\WittenNV .

One can change the value of the level to $k=h/2+n$ by wrapping $n$
five-branes\foot{Just as in the case discussed in \mnas , this
analysis is self-consistent when $n<<h$.}
 over ${\widetilde{\bf RP}}^3\in {\bf M}_7$, which in turn adds $n$
units of flux through ${\bf S}^3$. 
When $n>0$, adding
five-branes does not
break any further supersymmetry  and therefore supersymmetry is unbroken.
 On the other hand, when
$n<0$, anti-five-branes must be added to the background, which break
supersymmetry 
completely. Therefore, using SUGRA we have exactly reproduced the
conjecture by Witten regarding supersymmetry breaking as a function of
$k$ for these theories, that is for $k\geq h/2$ supersymmetry is
unbroken while supersymmetry is spontaneously broken for $k<h/2$.

The low energy effective theory when $n>0$ or $k>h/2$ is
properly described by the field theory on the $n$ wrapped branes over
${\widetilde{\bf RP}}^3$. The statement that the five-branes wrap
${\widetilde{\bf RP}}^3$ can be formalized by saying that there is a
map $\Phi$ between $Y$ and ${\widetilde{\bf RP}}^3$, where $Y$ is a
closed three-manifold. Since the background we are considering is
orientifolded, this means that in going around a noncontractible loop
in the target space, that the orientation of the worldsheet must be
reversed, which means that $\Phi$ must satisfy
\WittenXY\ $\Phi^*(x)=w_1(Y)$,
where $x\in H^1({\widetilde{\bf RP}}^3,Z_2)$ and $w_1(Y)$ is the
obstruction to the orientability of $Y$. Following \WittenXY ,
one can show that $\Phi$ must a map of even degree so we take, for
example,  $Y={\widetilde{\bf S}}^3$. By integrating the appropriate
Wess-Zumino coupling on the worldvolume of the $n$ wrapped
five-branes, 
we find  that
the flux over $Y$ is twice as large as that over ${\widetilde{\bf
RP}}^3$. Therefore, the effective field theory on the branes in the 
limit  where $n<<h$ in which
backreaction can be ignored is given by

\noindent
$\bullet\ G=SO(N):$ SO(n) Chern-Simons at level $k=N$

\noindent
$\bullet\ G= Sp(N):$ Sp(n) Chern-Simons at level $k=N$

This is to be compared with the expected low energy description found
by Witten and described in section 2

\noindent
$\bullet\ G=SO(N):$ SO(N) Chern-Simons at level $k_{eff}=n$

\noindent
$\bullet\ G= Sp(N):$ Sp(N) Chern-Simons at level $k_{eff}=n$

\noindent
Remarkably, the effective description predicted by the string dual is
precisely the level-rank dual \levelrank\ of the expected field theory
result, which has the same number of ground states.



\newsec{SUGRA dual to $D=4$ ${\cal N}=1$ SYM}

Four dimensional ${\cal N}=1$ SYM with gauge
group $G$ has at the classical level a chiral $U(1)_R$ symmetry that
acts on the gluino fields by   
\eqn\actonglu{
\lambda\rightarrow e^{-i\alpha} \lambda,\quad \bar{\lambda}\rightarrow
e^{i\alpha} \bar{\lambda}.} 
Likewise, $U(1)_R$ acts with the same charges on the supersymetry
generators  of positive
and negative chirality $Q_\alpha$ and $\bar{Q}_{{\dot \alpha}}$
respectively. Quantum mechanically, the presence of zero modes of
$\lambda$ in an instanton background induces a 't Hooft vertex which
explicitly breaks the $U(1)_R$ symmetry down to an anomaly free chiral
$Z_{2h}$ subgroup, where $h$ is the dual Coxeter number of $G$. The
$Z_{2h}$ symmetry is then spontaneously broken to $Z_2$, which acts
by $-1$ on $\lambda$ and $\bar{\lambda}$, by the presence of a gluino
condensate 
\eqn\glucon{
<\lambda\lambda>^h=\Lambda_{QCD}^{3h},}
which gives rise to $h$ supersymmetric vacua with a mass gap and
confinement.

\subsec{Supergravity Dual For $G=SU(N)$}

A simple way of realizing $D=4$ ${\cal N}=1$ $SU(N)$ SYM is to
consider the low energy effective field theory on $N$ five-branes
wrapping the ${\bf S}^2$ of the resolved conifold geometry, which is
topologically the $X={\cal O}(-1)\oplus {\cal O}(-1)$ bundle over ${\bf
S}^2$ and admits a metric with $SU(3)$ holonomy\foot{There is a
``mirror'' realization, in which pure SYM appears by wrapping $N$
D6-branes on the deformed conifold. It is in this Type IIA picture
where one can make a direct connection with M-theory on $G_2$
holonomy spaces
\AcharyaGB\AtiyahZZ .}. 

Recently, Maldacena and Nu\~nez \malnu\ have interpreted a gauged
supergravity solution 
found by Chamseddine 
and Volkov \ChamseddineNM\ChamseddineMC\ as the supergravity
description of the above wrapped five-brane configuration. This smooth
supergravity solution is thus dual to $D=4$ ${\cal N}=1$ $SU(N)$ SYM.

The solution they found is a compactification\foot{The internal
geometry is nevertheless non-compact.} of Type IIB string theory on
${\bf R}^{1,3}\times {\bf M}_6$ with a non-trivial three-form field
strength $H$ and dilaton over ${\bf M}_6$. The metric of the solution
is given by
\eqn\metricmal{\eqalign{
ds^2&=ds^2({\bf R}^{1,3})+N[dr^2+e^{2h(r)}(d\theta^2+\sin^2\theta\
d\phi^2)]+N{1\over 4}(\widetilde{w}_a-A_a)^2,\cr
A_1&=a(r)d\theta\qquad A_2=-a(r)\sin\theta d\phi\qquad A_3=\cos\theta d\phi}} 
where $\theta$ and $\phi$ parametrize the ${\bf S}^2$ which the branes
wrapped and $\widetilde{w}_a$ are $SU(2)$ left invariant one-forms parametrizing
the transverse 
${\widetilde{\bf S}}^3$ to the five-branes
\eqn\leftinv{\eqalign{
w_1&=\cos\widetilde\psi\
d\widetilde\theta+\sin\widetilde\psi\sin\widetilde\theta\
d\widetilde\phi\cr
w_2&=-\sin\widetilde\psi\
d\widetilde\theta+\cos\widetilde\psi\sin\widetilde\theta\
d\widetilde\phi\cr
w_3&=d\widetilde\psi+\cos\widetilde\theta\ d\widetilde\phi.}}
The explicit form of the radial functions $h(r),a(r)$ can be found in
\malnu . Here we will only need its asymptotics which near
$r\rightarrow \infty$ are $e^{2h}\simeq r$ and $a(r)\simeq 0$ while
near $r\simeq 0$ they behave as $e^{2h}\simeq r^2$ and $a(r)\simeq
1$ so that the metric in the square brackets in \metricmal\ is the
flat ${\bf R}^3$ metric near the origin. Topologically, ${\bf M}_6$
is asymptotically conical with base ${\bf S}^2\times {\widetilde{\bf
S}}^3$ and in the interior ${\widetilde{\bf
S}}^3$ stays of finite size while ${\bf S}^2$ becomes
contractible. The solution has ${\cal N}=1$ supersymmetry in four
dimensions so that ${\bf M}_6$ is a $SU(3)$ holonomy manifold with
respect to the connection with torsion $\bar\nabla=\nabla+\pi H$,
where $\nabla$ is the usual Levi-Civita connection\foot{The complex
geometry of ${\bf M}_6$ and the derivation of the BPS equations via
an effective superpotential for the ansatz can be found in
\PapadopoulosGJ.}
 while $H$ is the
three-form flux through ${\bf M}_6$.

The solution also has the following non-trivial three-form field strength
\eqn\Hsoln{
H={N\over 2\pi}\left[- {1\over 4}(w_1-A_1)\wedge (w_2-A_2)\wedge
(w_3-A_3)+{1\over 4}F_a\wedge ({\widetilde w}_a-A_a)\right]} 
whose flux over ${\widetilde{\bf
S}}^3$ encodes number of five-branes wrapping ${\bf S}^2$ in the 
 $X={\cal O}(-1)\oplus {\cal O}(-1)$ bundle over ${\bf
S}^2$
\eqn\fluxmalnun{
\int_{{\widetilde{\bf
S}}^3} {H\over 2\pi}=N.}

We note that this solution also realizes\foot{We note that a very
interesting description of a confining gauge theory in which branes
are replaced by fluxes also appeared in \KlebanovHB\ which is dual to a
cascading gauge theory.} the general philosophy of
\VafaWI . Here the original geometry of $X$ is such that it is
asymptotically a cone over ${\bf S}^2\times {\widetilde{\bf
S}}^3$  and five-branes wrap the topologically non-trivial ${\bf S}^2$ 
while ${\widetilde{\bf
S}}^3$ is contractible in the full geometry. In the dual supergravity solution,
the roles of ${\bf S}^2$ and ${\widetilde{\bf
S}}^3$ are reversed since in ${\bf M}_6$  ${\bf S}^2$ is contractible while ${\widetilde{\bf
S}}^3$ supports the flux left behind by the branes once they
disappear. The topology of ${\bf M}_6$ is that of the deformed
conifold. 

One can use this supergravity solution, which manifestly exhibits a
mass gap and confinement, to explore the gravity realization \malnu\ of the
expected chiral symmetry breaking $Z_{2N}\rightarrow Z_2$.
From the implementation of the twist required to obtain a
supersymmetric gauge theory on the wrapped brane one can identify the
classical $U(1)_R$ symmetry of the gauge theory with shift symmetry of
$\widetilde\psi$, which acts by $\widetilde\psi \rightarrow
\widetilde\psi +c$. Clearly, from the form of metric \metricmal\ in
the interior (IR) of ${\bf M}_6$ we see that the metric is only invariant
under shifts of $\widetilde\psi$ by $2\pi$ $\widetilde\psi \rightarrow
\widetilde\psi +2\pi$, which due to the  $\widetilde\psi$  periodicity
$\widetilde\psi \simeq
\widetilde\psi +4\pi$ generates a $Z_2$ symmetry, which is identified
with the $Z_2$ symmetry of ${\cal N}=1$ SYM preserved by the gluino condensate.

In the UV region of the geometry the full $U(1)_R$ symmetry acting by
arbitrary shifts on $\widetilde\psi$ is restored. On the other hand the
field theory expectation is that this symmetry is broken down to
$Z_{2N}$. In field theory, this explicit breaking is due to Yang-Mills
instantons and in \malnu\ this expectation was realized by taking into
account 
the effect of a Yang-Mills
instanton in the brane realization. By taking into account the
Wess-Zumino term \wesszumino\ 
one can indentify a SYM instanton
with  a string worldsheet wrapping an ${\bf S}^2$ in ${\bf M}_6$. As explained
in \malnu\ the appropiate description of a SYM instanton is given by
a string worldsheet wrapping an ${\bf S}^2$ in ${\bf M}_6$
parametrized by $\theta$ and $\phi$ and with $\widetilde\theta=\theta$
and  $\widetilde\phi=\phi$. Then the amplitude for a string instanton
wrapping this cycle is given by
\eqn\instan{
v=\exp\ \left(  i\int_{{\bf S}^2}B\right).}
By using the asymptotic expression for $H$ in \Hsoln\ one finds
that
\eqn\fase{
\int_{{\bf S}^2}B=-2N\widetilde\psi.}
Therefore, the instanton corrected UV string solution is only
invariant under $\widetilde\psi\rightarrow \widetilde\psi+ 2\pi k/N$,
where $n=1,\ldots, 2N$ and therefore reproduce  the expected anomaly
free $Z_{2N}$ R-symmetry.

In this way the expected chiral symmetry breaking pattern
$Z_{2N}\rightarrow Z_2$ is realized in the  string dual as the
symmetry group of the string solution being broken from the UV to the
IR. The $N$ vacua can then be easily constructed \malnu\ by performing
a discrete set of $N$  rotations of the $SU(2)$ gauge fields $A_a$ in
\metricmal\ 
described in \malnu\  such that
the UV solution is unchanged while it is rotated but still smooth in
the interior . We 
now extend this 
realization to SO and Sp gauge groups.

\newsec{Supergravity Dual For $G=SO$ and $G=Sp$}

One can generalize the construction in the previous section to $SO$
and $Sp$ gauge groups by introducing an ${\bf O}5$-plane on top of the
five-branes\foot{In the previous section we described the solution
with NS5-branes but can obtain the corresponding one for D5-branes by
performing an S-duality.}. In this case the  ${\bf O}5$-plane must
along ${\bf R}^{1,3}\times {\bf S}^2$, where ${\bf S}^2$ is the
supersymmetric cycle of the resolved conifold geometry $X={\cal
O}(-1)\oplus {\cal O}(-1)$ bundle over ${\bf 
S}^2$. This can be accomplished by considering
\eqn\modeloorin{ 
\hbox{Type IIB on}\ \left({\bf R}^{1,2}\times {\bf
X}\right)/\{1,\Omega\cdot \sigma\},}  
where $\Omega$ is the usual worldsheet parity operator and $\sigma$ is
$Z_2$ symmetry of  ${\bf X}$ whose fixed point set is ${\bf
R}^{1,3}\times {\bf S}^2$.

We recall that the conifold singularity
\eqn\conising{
z_1z_2-z_3z_4=0}
can be resolved by replacing \conising\ by  the following pair of
equations
\eqn\resol{
\pmatrix{z_1& z_3\cr\
z_4&z_2}\pmatrix{\lambda_1\cr
\lambda_2}=0}
where $(\lambda_1,\lambda_2)\simeq \lambda\ (\lambda_1,\lambda_2) $
don't vanish simultaneously and therefore parametrize an ${\bf
S}^2$. The system  
\resol\ is the familiar small resolution of the conifold singularity
and is topologically $X={\cal
O}(-1)\oplus {\cal O}(-1)$ bundle over ${\bf 
S}^2$. When $z_i=0\ \forall i$ \resol\ defines an entire ${\bf
S}^2$ and therefore resolves the conical singularity at the appex
while when any $z_i\neq 0$ \resol\ determines a unique point in ${\bf
S}^2$. $X$ is asymptotically conical with base ${\bf S}^2\times
{\widetilde{\bf S}}^3$ but ${\widetilde{\bf S}}^3$ is contractible.
With this description of $X$ it is easy to identify the $Z_2$
involution we were after, it is given by
\eqn\action{
{\sigma}:z_i\rightarrow -z_i\qquad i=1,\ldots,4.}
so that the fixed point set is $z_i=0$ where the ${\bf S}^2$
sits. Therefore, the orientifold \modeloorin\ results in an 
 ${\bf O}5$-plane along ${\bf
R}^{1,3}\times {\bf S}^2$.

We note for future reference that the singularity \conising\ has an
alternative desingularization, in which the singularity is deformed as
\eqn\defcon{
z_1z_2-z_3z_4=V.}
The action of $\sigma$ on this geometry is now free since the fixed
point set $z_i=0$ is not part of the geometry, so no orientifold plane
appears in this case.

The Types of gauge groups
that one can obtain by performing this orientifold are given in
Table 2. Likewise, the total five-brane charge -- that of the
five-branes plus that of the ${\bf O}5$-plane -- is given in Table 3.

Just like in the discussion in section $3$ the corresponding string
duals for these other gauge groups are obtained by implementing the
orientifold action above on the supergravity solution described in the
last section. Therefore, the string dual is given by an orientifold by
$\sigma$ of Type IIB on ${\bf R}^{1,3}\times {\bf M}_6$. ${\bf M}_6$ is
topologically like the deformed conifold geometry of \defcon\ so that
in this case the action of $\sigma$ is free and one obtains a dual
given by unoriented closed string theory on the regular geometry ${\bf
R}^{1,3}\times {\bf M}_6/Z_2$, without any open strings nor orientifold
planes. In this case the action of $\sigma$ is such that the
non-contractible ${\widetilde{\bf S}}^3$ inside ${\bf M}_7$ is
replaced\foot{This follows from the following observation. The real
section of equation \defcon\ describes ${\widetilde{\bf S}}^3$  so
that the effect of the $Z_2$ transformation $z_i\rightarrow -z_i$ is 
to replace it by ${\widetilde{\bf RP}}^3$. Topologically, the space is
$T^*{\widetilde{\bf RP}}^3$.}
by ${\widetilde{\bf RP}}^3$.

Apart from replacing ${\widetilde{\bf S}}^3\rightarrow {\widetilde{\bf
RP}}^3$ in ${\bf M}_6$ one must take into account the total five-brane
charge. Just like in section $3$ the total charge due to the branes
and the ${\bf O}5$-plane that we have on $X$ is replaced by flux over
the non-contractible ${\widetilde{\bf
RP}}^3$ of ${\bf M}_6$ so that we can replace in the solution found in
\malnu\ $N\rightarrow Q$, where $Q$ is the total five-brane charge
given in Table 3.

We can understand the expected pattern of chiral symmetry
breaking for these gauge theories by studying the symmetries of the
string solution in the UV and in the IR. In the IR, the metric looks
the same modulo the replacement ${\widetilde{\bf S}}^3\rightarrow
{\widetilde{\bf 
RP}}^3$  which makes the $\widetilde\psi$ coordinate $2\pi$ periodic,
that is $\widetilde\psi\simeq \widetilde\psi+2\pi$. Just as in the
previous section, the symmetry in the IR is just that given by shifts
by $2\pi$ $\widetilde\psi\simeq \widetilde\psi+2\pi$. This shift acts
as a trivial rotation in space-time but since the Killing spinors of
the supergravity solution are unmodified by the orientifold projection,
a rotation  by $2\pi$  on $\widetilde\psi$ generates a $Z_2$
transformation on the Killins spinors which is the expected unbroken
symmetry group after gluino condensation.

What is the symmetry in the UV? The symmetry of the metric is again 
$U(1)_R$ acting by arbitrary shifts on $\widetilde\psi$. We must analyze
the effect on the $U(1)_R$ symmetry of a SYM instanton in the string
theory realization. This can be accomplished by taking into account
the way a SYM instanton is represented in terms of branes in string
theory. Care must be taken in perfoming this identification since as
we discussed in section $3$ there are subtelties in the relation
between SYM instantons and D-branes for unoriented open strings. We
 showed that for $SO(N)$ 
gauge groups, via Wess-Zumino coupling \wesszumino , that a D1-string
on the D5-brane worlvolume corresponds to a single $SO(N)$ SYM
instanton. Therefore, the SYM instanton is represented by a euclidean
D1-string wrapping the ${\bf S}^2\in {\bf M}_6/Z_2$ that we described
in the previous section. The amplitude for such  a wrapped D1-string
is 
given by
\eqn\instan{
v=\exp\ \left(  i\int_{{\bf S}^2}B\right).}
By using the asymptotic expression for $H$ in \Hsoln\ and the required
replacement of $N\rightarrow Q$
one finds
that
\eqn\fase{
\int_{{\bf S}^2}B=-2Q\widetilde\psi.}
Therefore, the discrete rotations on $\widetilde\psi$ that act
non-trivially on the Killing spinors are given by
$\widetilde\psi\rightarrow \widetilde\psi +2\pi k/Q$ with
$k=1,\ldots,2Q$. By using Tables $3$ and $1$ we see that from the string
dual that the UV symmetry of the solution is precisely $Z_{2h}$ as
required by the field theory analysis.

For the $Sp(N)$ case there is an interesting subtlelty. As we
discussed in section $3$ the relation between brane charge and
instanton number is modified for $Sp(N)$ gauge theories. For an
$Sp(N)$ gauge theory on a collection of D5-branes we established that
a D1-string dissolved on the D5-branes actually corresponds to ${\it
two}$ instantons in the $Sp(N)$ gauge theory. Therefore, the amplitude
of a single SYM instanton is given by $v^{1/2}$ where $v$ is given by
\instan . The discrete rotations on $\widetilde\psi$ that leave
$v^{1/2}$ invariant and act
non-trivially on the Killing spinors are given by
$\widetilde\psi\rightarrow \widetilde\psi +2\pi k/Q$ with 
$k=1,\ldots,Q$. By taking the expression  for $Q$ in Table $3$
corresponding to the $Sp(N)$ case we see that we exactly reproduce the
expected  $Z_{2h}$ symmetry as can be seen from Table 1. 

Therefore, we have realized the
expected pattern of chiral symmetry breaking for $SO$ gauge groups
$Z_{2h}\rightarrow Z_2$ as the breaking of the UV symmetry of the
string solution to that of the IR. The $h$ vacua, just like in the
$SU(N)$ case, are obtained by performing the $h$ allowed discrete
rotations on  the 
$SU(2)$ gauge fields $A_a$ of  the
solution such that the new solutions are nonsingular in the interior 
but are equivalent in the UV region.

\medskip

\centerline{\bf Acknowledgments}

We would like to thank A. Brandhuber, E. Diaconescu, A. Kapustin,
J. Maldacena and 
E. Witten for discussions. The work of
J.G. is supported in part by the National Science Foundation under
grant No. PHY99-07949 and by the DOE under grant No. DE-FG03-92ER40701.

\vfill\eject

\appendix{A}{NS5 branes with $SU(2)_L$ gauge fields for 2+1 SYM}

Here we rederive the solution found by Maldacena and Nastase by
obtaining the first order BPS equations from an effective
quantum mechanical Lagrangian derived from gauged
supergravity. The relevant gauged supergravity is seven dimensional
$SO(4)$ gauged supergravity, which 
appears by reducing Type IIB supergravity on the near horizon geometry
of a NS5-brane. Since we are interested in finding a supergravity dual
to $D=3$ ${\cal N}=1$ SYM, the supersymmetry of the field theory on
the wrapped branes requires considering an $SU(2)_L$ truncation of 
$SO(4)$ gauged supergravity\foot{One can also have an 
$SU(2)_D=(SU(2)_L\times SU(2)_R)_{diag}$
truncation. Then , the solution has double the amount of
supersymmetry. This truncation was used in 
\GomisAA\GauntlettUR\   to find the
supergravity dual to D=3 ${\cal N}=2$ SYM.}.

The ansatz for  the solution is determined by the symmetries of the
background, which is given by five-branes wrapping ${\bf S}^3\in S({\bf
S}^3)$. This requires considering the most general ansatz invariant
under the $SO(4)$ isometries of ${\bf S}^3$.
In the string frame, the ansatz is therefore
\eqn\ansatz{\eqalign{
ds^2&=ds^2({\bf R}^{1,2})+dr^2+{e^{2h(r)}\over 4}w_aw_a\cr
A_a&={1+b(r)\over 2}w_a,}}
where $w_a, a=1,2,3$ are a set of $SU(2)$ left invariant one-forms satisfying 
\eqn\alge{
dw_a=-{1\over 2}\epsilon_{abc}\ w_b\wedge w_c.}
The three-form field strength $h$ of  $SU(2)_L$ gauged supergravity
satisfies the following Bianchi identify\foot{We are writing the
action such that the flux of $h/2\pi$ is quantized in integral units.}
\eqn\bianchi{
dh={1\over 8\pi}F_a\wedge F_a.}
The $SO(4)$ symmetry together with 
\eqn\field{
F_a={{\dot b}\over 2}dr\wedge w_a+{1\over
8}(b^2-1)\ \epsilon_{abc}w_b\wedge w_c}
determine the three-form from \bianchi \foot{The constant piece is
only fixed by  
requiring the solution we will find to be regular.}
\eqn\tres{
h={1\over 32 \pi}(b^3-3b+2)\ w_1\wedge w_2\wedge w_3.}
Plugging the ansatz into the bosonic $SU(2)_L$ gauged supergravity Lagrangian 
\eqn\lagran{
{\cal L}=\sqrt{g}e^{-2\phi}\left(R+4\partial_\mu \phi\partial^\mu \phi
-{1\over 8}F_{\mu\nu}^aF^{\mu\nu}_a-{\pi^2\over 3}
h_{\mu\nu\rho}h^{\mu\nu\rho}+4\right)}
 one obtains the following
effective quantum mechanical system
\eqn\effect{\
{\cal L}_{eff}=e^{2\alpha}(T-V),}
where
\eqn\kinpot{\eqalign{
T&=4{\dot \alpha}^2-3{\dot h}^2-{3\over 4}e^{-2h}{\dot b}^2\cr
V&={1\over 8}e^{-6h}(b^3-3b+2)^2+{3\over
4}e^{-4h}(b^2-1)^2-6e^{-2h}-4,}}
with
\eqn\definico{
\alpha={3\over 2}h-\phi\qquad {\dot \alpha}={d\alpha\over dr},\ldots .}
The $SU(2)_L$ gauged supergravity equations of motion follow from
\effect\ together with the Hamiltonian constraint 
which follows from  having fixed the
radial reparametrization invariance of the ansatz \ansatz 
\eqn\const{
H=T+V=0.}

One can find a set of first order BPS equations whose solutions also
solve the equations of motion if the potential of the effective
quantum mechanics can be written in terms of a superpotential as
follows
\eqn\superpot{
V={1\over 12}(\partial_hW)^2+{1\over 3}e^{2h}(\partial_bW)^2-{1\over
4}W^2.} The required superpotential is given by
\eqn\superpotans{
W=6e^{-h}\sqrt{M},}
where
\eqn\Mfin{
M=\left({1\over 12}e^{-2h}(b^3-3b+2)-b\right)^2+{1\over
4}e^{-2h}(b^2-1)^2-{2\over 3}(b^2-1)+{4\over 9} e^{2h}.}
The BPS equations following from this superpotential are given
by\foot{There is a typo in the Appendix of \mnas , the second
equation in $(4.4)$ should an extra multiplicative factor of $1/2$ in
the second term.} 
\eqn\BPSEQ{\eqalign{
{\dot \alpha}&={1\over 4}W= {3\over 2}e^{-h}\sqrt{M}\cr
{\dot b}&=-{2\over 3}e^{2h}\partial_b W={2\over
3\sqrt{M}}e^h\left[{1\over
8}e^{-4h}(b^3-3b+2)(1-b^2)+(1-b^3)e^{-2h}-2b\right]\cr
{\dot h}&=-{1\over 6}\partial_h W={e^{-h}\over
3\sqrt{M}}\left[e^{-4h}{(b^3-3b+2)^2 \over 16}+{e^{-2h}\over
2}(3(b^2-1)^2-2(b^4-3b^2+2b))+b^2+2\right].}} 
The asymptotics of the solution to these equations can be found in
\mnas .

This gauged supergravity solution can then be lifted to a solution of
Type IIB supergravity using the formulae in 
\CveticDM  . In the string frame
the Type IIB solution is 
\eqn\solun{\eqalign{
ds^2&=ds^2({\bf R}^{1,2})+ N[dr^2+{1\over 4} e^{2h} w_a w_a]+  N
{1\over 4}({\tilde w}_a-A_a)^2\cr
H&=h+{N\over 2\pi} \left[-{1\over 4}({\widetilde w}_1-A_1)\wedge({\widetilde
w}_2-A_2)\wedge({\widetilde w}_3-A_3)+{1\over 4}F_a\wedge({\widetilde
w}_a-A_a)\right]}}
where ${\widetilde w}_a, a=1,2,3$ are another set of $SU(2)$ left
invariant one forms parameterizing ${\widetilde{\bf S}}^3$.

This metric is defined on ${\bf R}^{1,2}\times {\bf M}_7$ where ${\bf
M}_7$ is parametrized by $r,w_a$ and ${\tilde w}_a$. $w_a$ and
${\tilde w}_a$ parameterize respectively ${\bf S}^3$ and ${\widetilde
{\bf S}}^3$. As described in the text ${\bf M}_7$ is asymptotically
topologically a
cone over ${\bf S}^3\times{\widetilde{\bf S}}^3$ but ${\bf S}^3$ is
contractible in the full geometry while ${\widetilde{\bf S}}^3$ is
topologically non-trivial.

\listrefs

\end